\documentclass[12pt,a4paper]{article}
\usepackage{amsmath,amssymb,type1cm,booktabs,bm,braket}
\usepackage[dvips]{graphicx}
\usepackage{amsmath,amssymb}
\usepackage{comment}	

\allowdisplaybreaks[1]
\numberwithin{equation}{section}

\renewcommand{\thefootnote}{\fnsymbol{footnote}}

\def\Lim{\qopname\relax\@empty{lim}\limits}
\let\lim = \Lim

\def\sup{\mathop{\operator@font sup}\limits}
\def\inf{\mathop{\operator@font inf}\limits}
\def\max{\mathop{\operator@font max}\limits}
\def\min{\mathop{\operator@font min}\limits}
\def\prod{\mathop{\mathchoice{\textstyle\prod@}{\textstyle\prod@}{%
      \scriptstyle\prod@}{\scriptscriptstyle\prod@}}\limits}
\def\coprod{\mathop{\mathchoice{\textstyle\coprod@}{\textstyle\coprod@}{%
      \scriptstyle\coprod@}{\scriptscriptstyle\coprod@}}\limits}
\def\bigcap{\mathop{\mathchoice{\textstyle\bigcap@}{\textstyle\bigcap@}{%
      \scriptstyle\bigcap@}{\scriptscriptstyle\bigcap@}}\limits}
\def\bigcup{\mathop{\mathchoice{\textstyle\bigcup@}{\textstyle\bigcup@}{%
      \scriptstyle\bigcup@}{\scriptscriptstyle\bigcup@}}\limits}
\def\Int{\displaystyle\intop\ilimits@}
\def\bigoplus{\mathop{\mathchoice{\textstyle\bigoplus@}{%
      \textstyle\bigoplus@}{\scriptstyle\bigoplus@}{%
      \scriptscriptstyle\bigoplus@}}\limits}
\newcommand{\siml}[0]{\hspace{0.3em}\raisebox{0.4ex}
	{$<$}\hspace{-0.7em}\raisebox{-0.7ex}{{\footnotesize $\sim$}}\hspace{0.4em}}
\newcommand{\simg}[0]{\hspace{0.3em}\raisebox{0.4ex}
	{$>$}\hspace{-0.7em}\raisebox{-0.7ex}{{\footnotesize $\sim$}}\hspace{0.4em}}

\def\slashchar#1{\setbox0=\hbox{$#1$}	
\dimen0=\wd0				
\setbox1=\hbox{/} \dimen1=\wd1		
\ifdim\dimen0>\dimen1			
\rlap{\hbox to \dimen0{\hfil/\hfil}}	
#1					
\else 					
\rlap{\hbox to \dimen1{\hfil$#1$\hfil}}	
/					
\fi}
\begin{document}
\begin{titlepage}

 \begin{flushright}
 \end{flushright}

 \vspace{1ex}

 \begin{center}

  {\Large \bf Beauty is more attractive: Particle Production and Moduli trapping
  with Higher Dimensional Interaction}

  \vspace{3ex}

  {\large $^{ab}$Seishi Enomoto\footnote{e-mail: enomoto@eken.phys.nagoya-u.ac.jp},
  $^c$Satoshi Iida\footnote{e-mail: saiida@th.phys.nagoya-u.ac.jp},\\
  $^{ac}$Nobuhiro Maekawa\footnote{e-mail: maekawa@eken.phys.nagoya-u.ac.jp},
  $^{d}$Tomohiro Matsuda\footnote{e-mail: matsuda@sit.ac.jp}}
  
  \vspace{4ex}
  {\it$^a$ Kobayashi Maskawa Institute, Nagoya University, Nagoya 464-8602, Japan}
  \\
  {\it$^b$ Institute of Theoretical Physics, Faculty of Physics, University of Warsaw, Ho$\dot{z}$a 69, 00-681 Warsaw, Poland}
  \\
  {\it$^c$ Department of Physics, Nagoya University, Nagoya 464-8602, Japan}
  \\
  {\it$^d$ Laboratory of Physics, Saitama Institute of Technology, Saitama 369-0293, Japan}
  \vspace{6ex}

 \end{center}

\begin{abstract}
We study quantum effects on moduli dynamics arising from particle production near the enhanced symmetry point (ESP).
We focus on non-renormalizable couplings between the moduli field and the field that becomes light at the ESP.
Considering higher dimensional interaction, we find that particle production is significant in a large area, 
which is even larger than the area that is expected from a renormalizable interaction.
It is possible to find this possibility from a trivial adiabatic condition;
however the quantitative estimation of particle production and trapping of the field in motion are far from trivial.
In this paper we study particle production and trapping in detail, using both the analytical and numerical calculations,
to find a clear and intuitive result that supports trapping in a vast variety of theories.
Our study shows that trapping driven by a non-renormalizable interaction is possible.
This possibility has not been considered in previous works.
Some phenomenological models of particle physics will be mentioned to complement discussion.
\end{abstract}

\end{titlepage}

\renewcommand{\thefootnote}{\arabic{footnote}}
\setcounter{footnote}{0}

\section{Introduction}

Supersymmetric (SUSY) models are one of the most promising candidates of
the theory beyond the standard model (SM).
One of the characteristic features of the SUSY models is that they have
a number of light scalar fields called flat directions or moduli, which
describes deformations of the effective system.
Therefore, these models may have many quasi-degenerated vacua, which
could be stable or metastable.
Our Universe may arise from one of those vacua.
Since the vacuum expectation value (VEV) of the moduli determines the
low-energy effective theory, it is important to find the way how these
moduli find their vacuum when the moduli are not static.
Thinking about the very early stage of the cosmological evolution, these
fields might have a large kinetic energy compared to their potential
energy.

A decade ago, Kofman et.al. suggested in their paper ``Beauty is
attractive''~\cite{Kofman:2004yc} that a vacuum could be chosen because
it is attractive.
Their critical observation is that since new degrees of freedom become
light when the modulus passes through the enhanced symmetry point (ESP), these species could be
produced by the quantum effects and could alter the dynamics in such a
way as to drive the moduli towards the ESP.
They called this mechanism ``trapping''.
Their basic argument was based on the effective theory that contains an
interaction between two scalar fields:
\begin{equation}
\label{conventional-int}
 \mathcal{L} = \partial_{\mu} \phi^* \partial^{\mu} \phi + \frac{1}{2} \partial_{\mu} \chi \partial^{\mu} \chi
  - \frac{1}{2} g^2 \left| \phi \right|^{2} \chi^2.
\end{equation}
The interaction may arise in a system of moving branes,
in which a gauge symmetry of the effective action could be enhanced when
branes pass through.

On the other hand, we know that in some cases moduli in the effective
action may not be coupled to light fields through renormalizable
couplings, in the sense that they may have interactions suppressed by the
cutoff scale. 
A typical example could be the neutrino mass in the see-saw mechanism,
which couples to the Higgs scalar field through higher dimensional
interaction in the low energy effective theory.\footnote{
Of course the Higgs is not a moduli field. We think there is no
confusion in this analogy.}
Note that in this paper we are considering a more general situation,
which has not been considered in Ref.~\cite{Kofman:2004yc}. 
We are extending the original scenario to include higher dimensional
interaction, to show that such interaction is indeed responsible for
particle production and trapping.  We consider interaction:
\begin{equation}
 {\cal L}_\mathrm{int}\sim \frac{g^2\left| \phi \right|^{2n} \chi^2}{\Lambda^{2(n-1)}},
\end{equation}
which is sometimes called ``higher dimensional'' interaction, since
the mass dimension of $|\phi|^{2n}\chi^2$ is higher than
four.\footnote{Alternatively, particle creation and dissipation
could be caused by the  
Kaluza-Klein tower of the extra dimensions~\cite{Matsuda:2012kc}.
More recently, ``higher dimensional field space'' has been considered in
Ref.~\cite{Battefeld:2010sw, Battefeld:2013bfl}. We hope there is no confusion in those
``higher dimensional'' arguments.}  

One might think that higher dimensional interaction is not
important because it is suppressed by the cutoff scale; however a trivial
adiabatic condition ($|\dot{\omega}_k/\omega_k^2|>1$, which will be
defined later in this paper) suggests that particle production is
possible in a wider area than conventional trapping.
Although the qualitative argument based on the adiabatic condition
is suggesting that higher dimensional interaction could be important, 
it is not quite obvious whether such interaction is 
responsible for trapping.
Obviously, the quantitative estimation of quantum particle
production is far from trivial.

In this paper, we study the production of the particles when higher
dimensional interaction is responsible for the mechanism.
The amount of particle production will be calculated extending the
method developed in Ref.~\cite{Chung:1998bt}.
Our conclusion is that particle production via higher dimensional interaction
is significant in a wide area around the ESP.
Our result is consistent
with the naive estimation based on the adiabatic condition.
Moreover, using the numerical calculation, we demonstrate thattrapping
is indeed possible via higher dimensional interaction.
Interestingly, as is expected from the adiabatic condition, we confirmed
that the area of trapping becomes wider than conventional
trapping. 

Our conclusion is that the mechanism can play an important role in
selecting the vacuum in the early Universe.
Note that even if one is considering the conventional ESP, one might
have to consider higher dimensional interaction, since in reality
there could be multiple fields that are responsible for the symmetry
breaking.
In that case quantum particle production must be calculated when the
original symmetry is already broken by the other fields (but higher
dimensional interaction could remain).\footnote{We need a
field $\chi_{nc}$ whose mass is determined by higher
dimensional interaction. Note that $\chi_{nc}$ could not be identical to the
conventional $\chi_c$, which becomes massive when the gauge symmetry is broken.
See Sec.\ref{sec:toy_model} for more details, in which 
we consider a toy model.
Particle production in the multi-field dynamics is rather
complicated.}
This possibility has never been considered before.
We found that trapping is indeed possible for such
ESPs, where the gauge symmetry could not be recovered but higher
dimensional interaction may arise.

\section{Basic mechanism of quantum particle production and trapping at an ESP}
 \label{sec:Review_of_Particle_Production_and_Vacuum_Selection}
First we review the basics of particle production and trapping of Ref.~\cite{Kofman:2004yc}.
Consider the Lagrangian:
\begin{equation}
 \mathcal{L} = \partial_{\mu} \phi^* \partial^{\mu} \phi + \frac{1}{2} \partial_{\mu} \chi \partial^{\mu} \chi
  - \frac{1}{2} g^2 \left| \phi \right|^{2} \chi^2, \label{eq:Lagrangian_basic}
\end{equation}
which consists of two scalar fields $\phi$ (complex) and $\chi$ (real).
Here, $\phi$ is a field in classical motion, while $\chi$ is the
 field that is produced by the quantum effect.
For simplicity, we assume that $\phi$ is homogeneous in space, so that we
can write $\phi=\phi(t)$.

When $\chi\simeq 0$ and the back-reaction is negligible, the equation of
motion gives 
\begin{equation}
 \phi (t) = vt + i \mu, \label{eq:WKB_solution_of_phi}
\end{equation}
where $v$ is the velocity of $\phi$, and $\mu$ is called the impact parameter.
This solution is valid as far as the adiabatic condition
($\dot{\omega}_k/\omega_k^2<1$) is satisfied for $\omega_k$.
Here, we defined 
\begin{equation}
 \omega_k=\sqrt{\mathbf{k}^2 + g^2 \left| \phi \right|^2},
\end{equation}
where $\mathbf{k}$ is the momentum of $\chi$.
For a low momentum mode ($\mathbf{k} \sim 0$) and a small impact
parameter ($\mu^2 \ll v$), the adiabatic condition leads to
\begin{equation}
 \left| \phi \right| \simg \sqrt{v/g}.
\end{equation}

\begin{figure}[t]
 \begin{center}
  \includegraphics[scale=0.7]{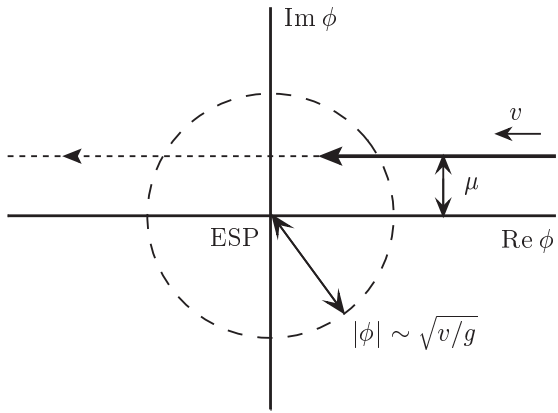}
  \caption{The trajectory of $\phi(t)$ is shown on the $\phi$-space.
$\chi$-particles are produced in the area inside the circle ($\left| \phi \right| \le \sqrt{v/g}$).}
  \label{fig:ESP_closing}
 \end{center}
\end{figure}

Consider the trajectory shown in Figure \ref{fig:ESP_closing}, in which $\phi$
comes into the ``non-adiabatic'' area $\left| \phi \right| \siml
\sqrt{v/g}$, where the adiabatic condition is violated and the quantum
back-reaction is not negligible.
The observation in Ref.~\cite{Kofman:2004yc} is that in that area the
mass of $\chi$ becomes light and the kinetic energy of $\phi$ can be
translated into $\chi$-particle production.
The number density $n_{\chi}$ of the field $\chi$ has been evaluated  as
\begin{equation}
 n_{\chi} = \frac{(gv)^{3/2}}{(2\pi)^3} e^{- \pi g \mu^2 / v}. \label{eq:number_density_for_n=1}
\end{equation}
This result shows that particle production is indeed significant
 in the area $|\phi| \siml \sqrt{v/g}$.
The result is consistent with the naive estimation from the adiabatic
condition.

\begin{figure}[t]
 \begin{center}
  \includegraphics[scale=0.5]{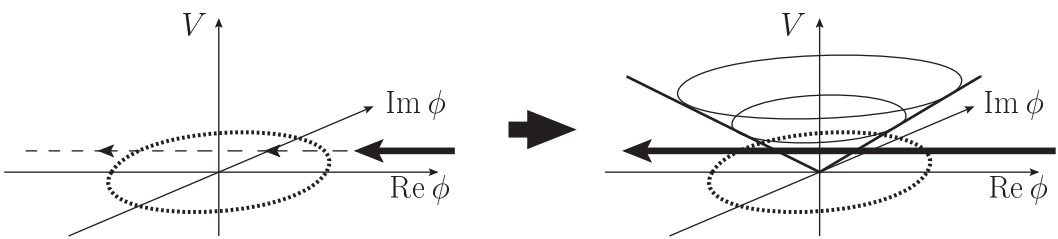}
  \caption{$\chi$-particles are produced when $\phi$ goes into the
  non-adiabatic area (inside the dotted circle in the left-hand side picture). 
  Quantum creation of the $\chi$-particles induces a linear potential
  for $\phi$. (See the picture on the right-hand side.)}
  \label{fig:established_potential}
 \end{center}
\end{figure}

\begin{figure}[ht]
 \begin{center}
  \includegraphics[scale=0.7]{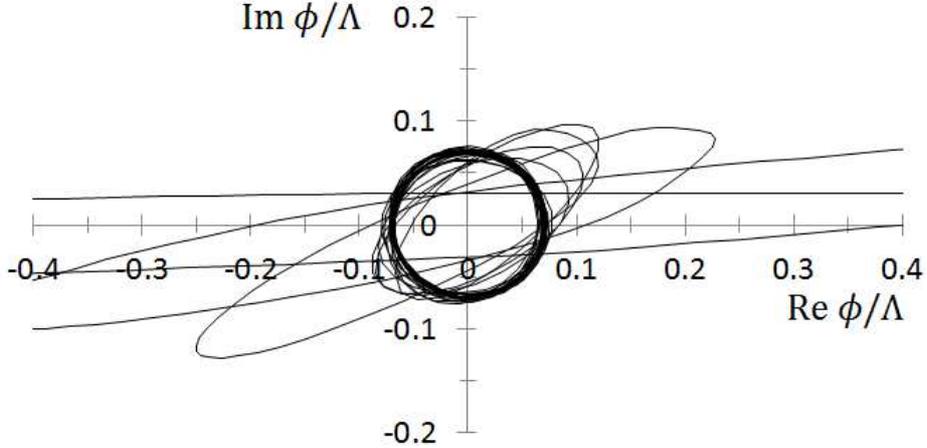}
  \caption{The solid line is the trajectory of $\phi$ for
$g^2=20, v=0.1$ and $\mu=0.03$.}
  \label{fig:trapping}
 \end{center}
\end{figure}

After the quantum excitation of $\chi$, $\phi$ moves away from the ESP.
However, the interaction causes back-reaction, since the motion 
increases the mass of $\chi$ particle ($m_{\chi}=g|\phi|$).
As a result, the energy density of $\chi$ increases as $\phi$ moves away from the
ESP. One will find
\begin{equation}
 \rho_{\chi} \sim m_{\chi} n_{\chi} = g|\phi|n_{\chi}.
\end{equation}
The energy density $\rho_\chi$ changes the potential of $\phi$.
The induced potential is a linear potential measuring the distance from the ESP.
See also Fig.~\ref{fig:established_potential}.
After that, $\phi$ goes up the potential until the initial kinetic
energy is comparable to the 
potential energy, where $\phi$ makes a turn and goes back to the ESP again.
Then, there will be another production of $\chi$, giving additional
 back-reaction.
Finally, $\phi$ is ``trapped'' around the ESP.
Numerical calculation of the trajectory is shown in Figure
\ref{fig:trapping}. 

\section{Analytical calculation with higher dimensional interaction}
In the previous section we reviewed quantum particle production when
$\phi$ passes through the non-adiabatic area around the ESP. 
Trapping has already been confirmed~\cite{Kofman:2004yc} for
a renormalizable interaction given in Eq.(\ref{eq:Lagrangian_basic}).
In this section, we consider higher dimensional interaction instead of
renormalizable one.
We start with the Lagrangian:
\begin{equation}
 \mathcal{L} = \partial_{\mu} \phi^* \partial^{\mu} \phi + \frac{1}{2} \partial_{\mu} \chi \partial^{\mu} \chi
  - \frac{1}{2} \frac{g^2}{\Lambda^{2(n-1)}} \left| \phi \right|^{2n} \chi^2, \label{eq:Lagrangian_with_HDI}
\end{equation}
where $n=1$ reproduces a renormalizable interaction.
One can check the consistency of our calculation for $n=1$.

First consider the non-adiabatic condition.
From $\dot{\omega}_k/\omega_k^2 \simg 1$, 
we obtain
\begin{equation}
 |\phi| \siml \sqrt{\frac{v}{g}} \cdot \left( \frac{g \Lambda^2}{v} \right)^{\frac{n-1}{2(n+1)}}, \label{eq:particle_production_area}
\end{equation}
where $\dot{\phi}=v$ and $|\phi + \phi^*| \sim 2|\phi|$ are considered.
The above (naive) condition shows that the area may become even larger
for larger $n$.
A similar speculation could arise from the effective mass near the ESP,
which is given by
\begin{equation}
 m_{\chi} = \frac{g}{\Lambda^{n-1}} \left| \phi \right|^{n}
\sim\frac{g}{\Lambda^{n-1}} \left| \mu \right|^{n}
 \label{eq:mass_of_chi_HDI}.
\end{equation}
One can see that $m_\chi$ is lighter for larger $n$.
Since the kinematically allowed mass ($m_\chi$) is bounded from above
by the initial kinetic energy of $\phi$, we can expect that the area of particle
production could be wider for larger $n$.
On the other hand, the interaction seems to be weaker for larger
$n$.
Therefore, there could be a tension between these effects, 
which cannot be solved without looking into more details of 
the mechanism.

\subsection{Equations and trajectory without particle production} \label{sec:solutions}
Before solving Eq.(\ref{eq:Lagrangian_with_HDI}), we need to include 
$\delta V$ for the normalization:
\begin{equation}
 \mathcal{L} = \partial_{\mu} \phi^* \partial^{\mu} \phi + \frac{1}{2} \partial_{\mu} \chi \partial^{\mu} \chi
  - \frac{1}{2} \frac{g^2}{\Lambda^{2(n-1)}} \left| \phi \right|^{2n} \chi^2 + \delta V, \label{eq:Lagrangian_with_HDI_and_ct}
\end{equation}
where $\delta V$ is defined by
\begin{equation}
 \delta V = \frac{1}{2} \int \frac{d^3 k}{(2 \pi)^{3/2}} \: \omega_k,
\end{equation}
and $\omega_k$ is a frequency of the quantum field $\chi$ defined by
\begin{equation}
 \omega_k=\sqrt{\mathbf{k}^2 + \frac{g^2}{\Lambda^{2(n-1)}}|\phi|^{2n}}. \label{eq:frequency}
\end{equation}
$\delta V$ is needed to subtract a divergence in the Coleman-Weinberg potential.
Note that we are considering the effective action that has an explicit cutoff scale.

Next, we define the wave function $u_{\mathbf{k}}$ for $\chi$.
Then the creation/annihilation operators $a_{\mathbf{k}}, a_{\mathbf{k}}^{\dagger}$
are defined as
\begin{equation}
 \chi (t, \mathbf{x}) = \int \frac{d^3 k}{(2\pi)^{3/2}} \: e^{i\mathbf{ k\cdot x}}
  \left( a_{\mathbf{k}} u_{\mathbf{k}}(t) + a_{\mathbf{-k}}^{\dagger} u_{\mathbf{-k}}^*(t) \right),
\end{equation}
where $a_{\mathbf{k}}, a_{\mathbf{k}}^{\dagger}$ satisfy the commutation relations
\begin{equation}
 \left[ a_{\mathbf{k}}, a_{\mathbf{k'}} \right] = \left[ a_{\mathbf{k}}^{\dagger}, a_{\mathbf{k'}}^{\dagger} \right] = 0,
  \quad \left[ a_{\mathbf{k}}, a_{\mathbf{k'}}^{\dagger} \right] = \delta^3(\mathbf{k-k'}),
\end{equation}
and $u_{\mathbf{k}}$ satisfies the norm condition
\begin{equation}
 \dot{u}_{\mathbf{k}}^* u_{\mathbf{k}} - u_{\mathbf{k}}^* \dot{u}_{\mathbf{k}} = i.
\end{equation}
We obtain the equations of motion:
\begin{equation}
 \displaystyle  \ddot{\phi} + \frac{ng^2 |\phi|^{2(n-1)}}{2 \Lambda^{2(n-1)}} \phi
  \int \frac{d^3 k}{(2\pi)^3} \left( |u_{\mathbf{k}}|^2 - \frac{1}{2\omega_k} \right) = 0, \label{eq:EOM_of_phi}
\end{equation}
\begin{equation}
 \ddot{u}_{\mathbf{k}} + \omega_k^2 u_{\mathbf{k}} = 0. \label{eq:EOM_of_u}
\end{equation}
Here, we assumed that $\phi(t)$ is homogeneous in space.
Since  $\omega_k$ is the homogeneous function of $\phi(t)$, 
the wave function $u_k = u_{\mathbf{k}} = u_{\mathbf{-k}}$ is
also homogeneous in space.

Let us examine the initial conditions. 
For $\phi(t)$, we can define initial quantities $\phi_0$ and $\dot\phi_0$ at
$t=-\infty$.
Using the WKB approximation for $u_k(t)$, we consider the initial wave function
\begin{equation}
 u_k(t) \sim \frac{1}{\sqrt{2\omega_k(t)}} \: e^{-i \int_{-\infty}^t dt'\omega_k(t')}. \label{eq:WKB_solution_of_u}
\end{equation}
This WKB solution is valid as far as the adiabatic condition is satisfied:
\begin{equation}
 \left| \frac{d}{dt} \frac{1}{\omega_k} \right| \ll 1. \label{eq:WKB_valid_region}
\end{equation}
Substituting (\ref{eq:WKB_solution_of_u}) into (\ref{eq:EOM_of_phi}), we
obtain the trivial solution 
\begin{equation}
 \phi(t) \sim vt + i\mu, \label{eq:WKB_solution_of_phi_HDI}
\end{equation}
which reproduces the classical motion (\ref{eq:WKB_solution_of_phi}).
The above solution is valid when quantum production and the back-reaction are negligible.

\subsection{Quantum particle production and the back-reaction} \label{sec:particlpe_production_and_established_potential}
As $\phi$ approaches toward the ESP, the adiabatic condition 
(\ref{eq:WKB_valid_region}) will be violated.
In that case the WKB approximation (\ref{eq:WKB_solution_of_u}) is
no longer valid.
We consider~\cite{Chung:1998bt,Birrell:1982ix}
\begin{equation}
 u_k(t) = \frac{\alpha_k(t)}{\sqrt{2\omega_k}} e^{-i\int_{-\infty}^t dt'\omega_k(t')}
   + \frac{\beta_k(t)}{\sqrt{2\omega_k}} e^{+i\int_{-\infty}^t dt'\omega_k(t')}. \label{eq:Bogoliubov_transformed_solution}
\end{equation}
The equation of motion (\ref{eq:EOM_of_u}) gives
\begin{eqnarray}
 \dot{\alpha}_k &=& \beta_k \frac{\dot{\omega}_k}{2\omega_k} e^{+2i\int_{-\infty}^t dt' \omega_k}\\
 \dot{\beta}_k &=& \alpha_k \frac{\dot{\omega}_k}{2\omega_k} e^{-2i\int_{-\infty}^t dt' \omega_k}, \label{eq:EOM_of_beta}
\end{eqnarray}
where the initial conditions are $\alpha_k(-\infty)=1$ and $\beta_k(-\infty)=0$.
We can evaluate the number density of the light field from
$\beta_k(\infty)$, using the relation
\begin{equation}
n_\chi=\int\frac{d^3k}{(2\pi)^3}|\beta_k|^2.
\end{equation}
Considering $\phi(t)=vt+i\mu$ as the 0-th order approximation, one can find 
the exact solution when $n=1$. (See the review in Sec.\ref{sec:Review_of_Particle_Production_and_Vacuum_Selection}.) 
The situation changes when $n\geq 2$, since we do not have the obvious
solution that can be used for $n_\chi$.
Therefore, we need another calculational method that 
is valid for higher dimensional interaction.
An obvious way is to find $n_\chi$ using the numerical calculation,
which will be demonstrated later in this paper.
The numerical calculation supports our analytical calculation.

Because of the complexity of the method, the details 
of the analytical calculation will be described in Appendix
\ref{sec:calculation_of_produced_particle_number}.  
We obtained 
\begin{equation}
 n_{\chi} =  C_n^{(0)} \cdot (gv)^\frac{3}{2} \left( \frac{v}{g\Lambda^2} \right)^{\frac{3(n-1)}{2(n+1)}}
  \left( 1 + C_n^{(1)} \cdot \left( \frac{g\Lambda^2}{v} \right)^{\frac{2}{n+1}} \frac{\mu^2}{\Lambda^2}
   + \mathcal{O} \left( \frac{\mu^4}{\Lambda^4}\right)  \right),
   \label{eq:number_density_formula}
\end{equation}
where $C_n^{(0)}$ and $C_n^{(1)}$ are numerical coefficients. 
(See Table \ref{tab:coefficients}.)
\begin{table}[b]
\begin{center}
 \begin{tabular}{|c||c|c|c|}
  \hline
 &$n=1$ & $n=2$ & $n=3$ \\ \hline \hline
  $C_n^{(0)}$ & $0.0044210$ & $0.012121$ & $0.035462$ \\ \hline
  $C_n^{(1)}$ & $-\pi\sim -3.1416$ & $-1.9251$ & $-0.64545$ \\ 
  \hline
 \end{tabular}
  \caption{Coefficients $C_n^{(0)}, C_n^{(1)}$ in case $n=1,2,3$.}
 \label{tab:coefficients}
\end{center}
\end{table}
This formula is valid for the impact parameter
\begin{equation}
 \mu \siml \left( \frac{v\Lambda^{n-1}}{g} \right)^{\frac{1}{n+1}}. \label{eq:valid_region}
\end{equation}
Considering $n=1$, one can easily check that Eq.(\ref{eq:number_density_formula}) is 
consistent with Eq.(\ref{eq:number_density_for_n=1}).
Note that the leading order (the first term in eq. (\ref{eq:number_density_formula})) 
is independent of the impact parameter $\mu$.
(Note however that the condition (\ref{eq:valid_region}) is crucial.)
Moreover, if we consider a large-$n$ limit,
the area of particle production becomes larger.  
Significantly, the area of quantum particle production can become
 larger than the conventional
scenario, although the number density ($n_\chi$) has a suppression
factor $\left(\frac{v}{g\Lambda^2}\right)^{\frac{3(n-1)}{2(n+1)}}$ when
$v \ll g\Lambda^2$. 
The above result is consistent with the naive argument based on the adiabatic condition. 
Of course, we need to examine the back-reaction from particle
production in more detail; otherwise trapping is not obvious.

The back-reaction is caused by the effective potential generated by the
energy density $\rho_\chi=m_\chi n_\chi$:
\begin{equation}
V_\mathrm{eff}(\phi)\sim \rho_{\chi} \sim \frac{g}{\Lambda^{n-1}}|\phi|^{n} n_{\chi}. \label{eq:additional_potential}
\end{equation}
Note that the Hamiltonian gives
\begin{equation}
 \left< H \right> = \int d^3x \left( |\dot{\phi}|^2 + \rho_{\chi} \right),
\end{equation}
where $\left< H \right>$ is the VEV of the Hamiltonian $H$.
Here, we can see that trapping force is weak near the ESP,
but it becomes stronger when $\phi$ is away from the ESP.
Except for a low-velocity limit ($v\ll g\Lambda^2$), trapping
force is significant when $\phi$ approaches $\phi\sim\Lambda$.

In the next section, we will examine the trapping mechanism 
using the numerical calculation.

\section{Numerical results}
In this section we present our numerical
calculation and compare it with the analytical estimation, so that one
can easily examine the consistency between them.
Then, we can examine the trapping mechanism using the trajectory $\phi(t)$.

Our numerical results are obtained by solving the equations of motion
(\ref{eq:EOM_of_phi}) and (\ref{eq:EOM_of_u}). 
The initial condition is defined using the WKB approximation, since
$\phi$ is initially away from the non-adiabatic area.
In this paper we consider $\phi(t_0)=5\Lambda+i\mu$, where the initial
velocity $\dot{\phi}(t_0)$ is considered for $v=-0.1\Lambda^2$ and $ -0.01\Lambda^2$.
For $u_k$, we take $u_k(t_0) = 1/\sqrt{2\omega_k}$, where
$\dot{u}_k(t_0)$ is obtained by taking the differential
of Eq.(\ref{eq:WKB_solution_of_u}), which gives
\begin{equation}
 \dot{u}_k(t_0)=-\left( \frac{\dot{\omega}_k}{2\omega_k} + i\omega_k \right) u_k(t_0).
\end{equation}

\subsection{Number density after quantum particle production}
\label{sec:produced_number_density}
\begin{figure}[t]
 \begin{center}
  \includegraphics[scale=0.5]{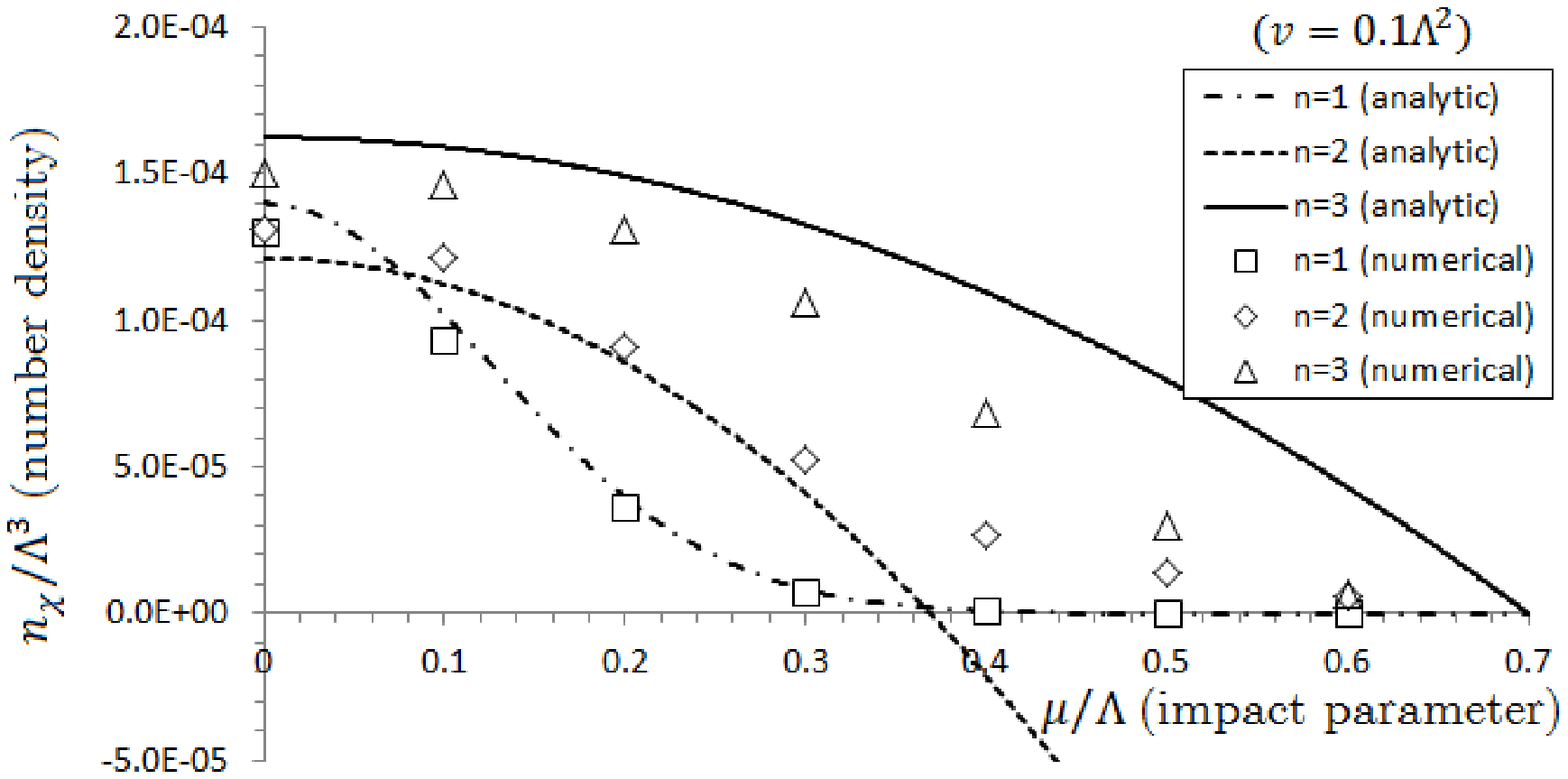}\\[2ex]
  \includegraphics[scale=0.5]{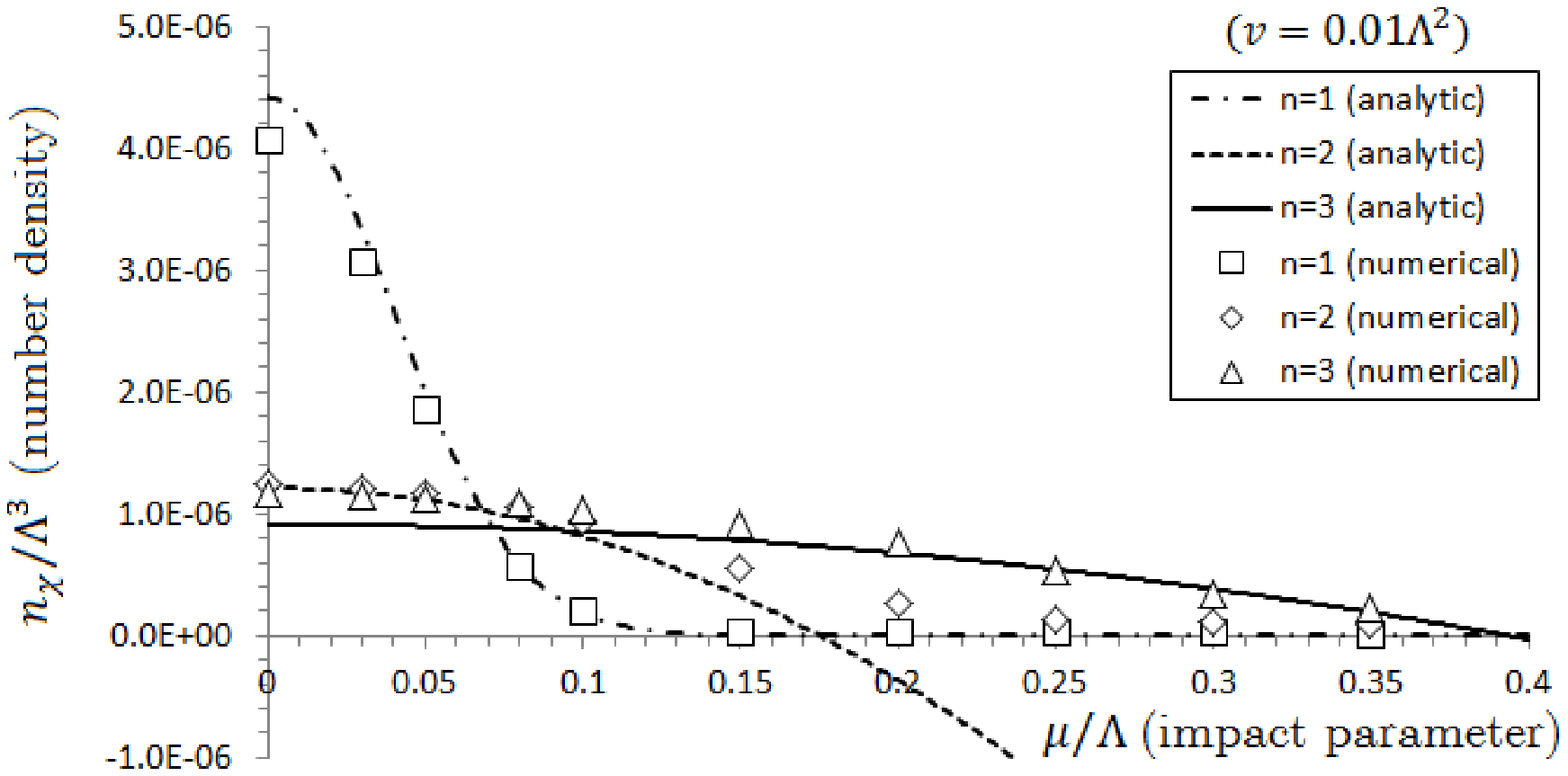}
  \caption{The horizontal axis ($\mu/\Lambda$) is the impact parameter,
and the vertical axis ($n_{\chi}/\Lambda^3$) is the number density,  both
  are normalized using a cutoff scale.  
From the vertical axis ($n_{\chi}/\Lambda^3$), we can see 
the normalized potential ($V(\phi)\sim m_\chi(\phi) n_\chi$) evaluated at 
$|\phi|\sim\Lambda$. 
The upper shows our result for $g=1$ and $\: v=-0.1\Lambda^2$,
   and the lower shows a result for $g=1$ and $\: v=-0.01\Lambda^2$.
Points marked with a triangle, square and rhombus are showing the numerical
  results. Lines are showing the analytical results.
Eq.(\ref{nchi_n1}) is used for the analytical solution
  of $n=1$, and Eq.(\ref{eq:number_density_formula}) is used for $n=2$
  and $3$.}
  \label{fig:number_density}
 \end{center}
\end{figure}

The number density after quantum particle production is obtained
from $u_k$ and $\dot{u}_k$~\cite{Garbrecht:2002pd}:
\begin{equation}
 n_{\chi} = \int \frac{d^3k}{(2\pi)^3} \left( \frac{|\dot{u}_k|^2+\omega_k^2|u_k|^2}{2\omega_k} - \frac{1}{2} \right).
\end{equation}
We show our results for $v=-0.1\Lambda^2$ and $v=-0.01\Lambda^2$ in
Figure \ref{fig:number_density}. 
Considering the initial energy density (for $v^2=0.01\Lambda^4$  and $10^{-4}\Lambda^4$), 
we can examine the ratio of the induced potential energy at $|\phi|=\Lambda$
to the initial kinetic energy.
Here, the vertical axis in figure 4 gives the potential energy at $|\phi|=\Lambda$. 
From the figure, we can see that the ratio is roughly 1/100 after the first event of
quantum particle production. 
The figure shows that our analytical estimation is in
good agreement with the numerical calculation.  
We can also confirm that the area for quantum particle production is
expanded for larger $n$.

\subsection{Time evolution and dynamics} \label{sec:time_evolution_of_dynamics}
In Sec.~\ref{sec:particlpe_production_and_established_potential},
the number density is calculated just for the first impact.
In reality, $\phi$ is pulled back toward the ESP, and the production
occurs many times.
Since the analytical calculation is not suitable for that purpose, 
we solved Eq.(\ref{eq:EOM_of_phi}) and (\ref{eq:EOM_of_u}) numerically
and examined the dynamics of trapping. 

\begin{figure}[t]
 \begin{center}
  \includegraphics[scale=0.5]{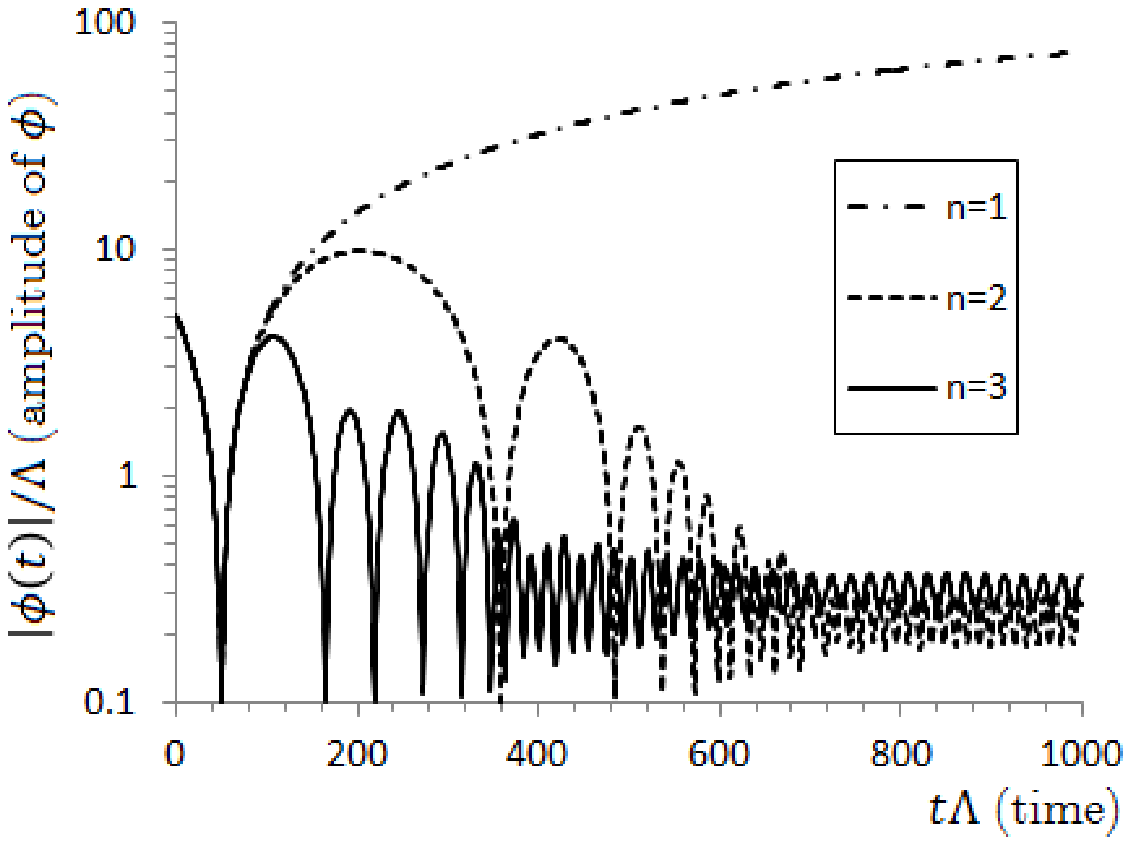}
  \includegraphics[scale=0.5]{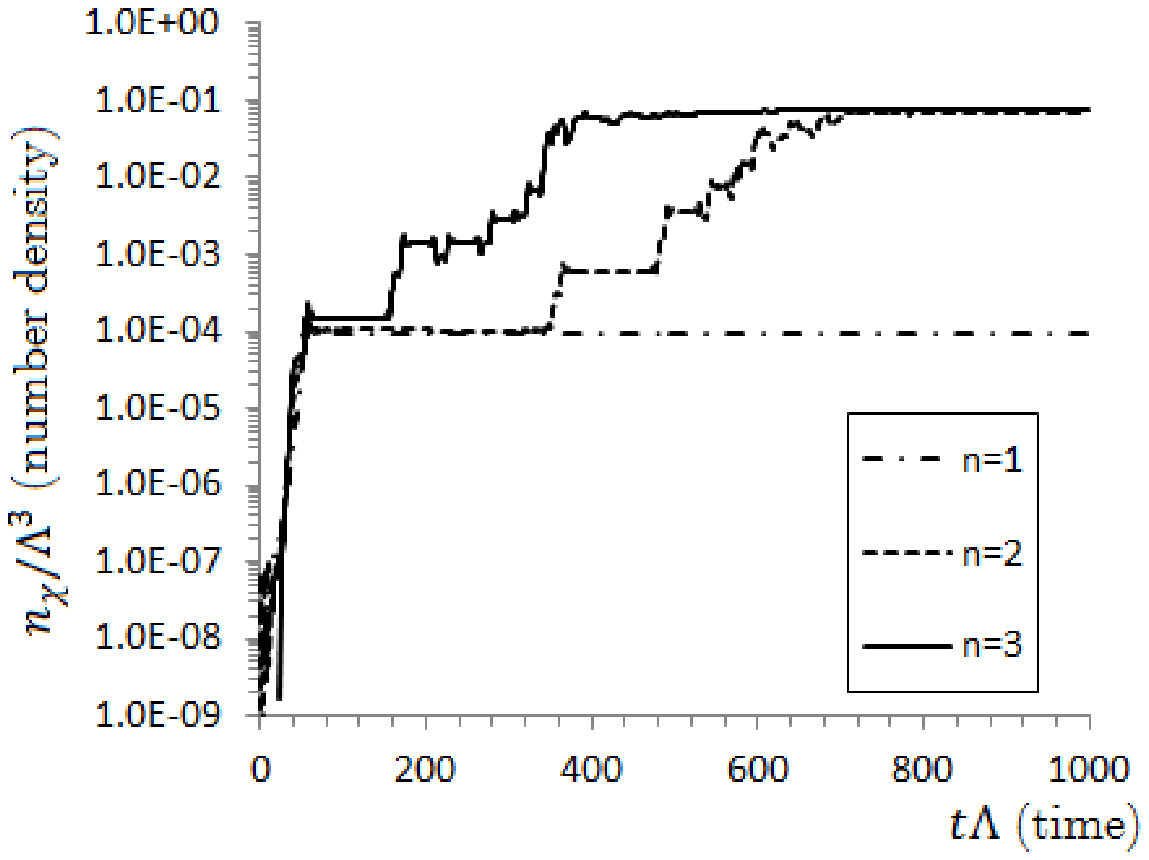}
  \caption{In both pictures the horizontal axes are the time
  ($t\Lambda$), which is normalized using $\Lambda$.
On the left-hand side, the vertical axis is $|\phi|/\Lambda$,
while on the right-hand side, the vertical axis is $n_{\chi}\Lambda^3$.
The time evolution toward trapping is shown on the left-hand side.
The produced number density $n_\chi/\Lambda^3$ for $g=1$, $
  v=0.1\Lambda^2,$ and $\mu=0.1\Lambda$ is shown on the right-hand side.   
We can see that trapping is quick for $n=3$, if it is
  compared with conventional trapping ($n=1$). 
For $n=1$, the typical time scale of trapping is beyond the scale of
  the figure.}
  \label{fig:amplitude}
 \end{center}
\end{figure}

We calculate the time evolution of $|\phi|$ and $n_{\chi}$ for
$g=1, v=0.1\Lambda^2, \: \mu=0.1\Lambda$, and $n=1,2,3$.
The results are shown in Figure \ref{fig:amplitude}, in which we can see
that trapping is quick for $n=3$. 

Although higher dimensional interaction becomes significant (i.e, the effective
potential becomes steep) for $|\phi|>\Lambda$, in reality the effective theory could
not be valid for $|\phi| > \Lambda$.
In that case, the back-reaction has to be explained using
 the original action (or by another effective action which has another cutoff).
In any case the calculation becomes highly model dependent and is not
suitable for our purpose.
Here, we have considered a modest assumption that higher dimensional interaction is
valid for $|\phi|< 10 \Lambda$.
Alternatively, one can avoid the problem by increasing the
total number of $\chi$ species to $\mathcal{O}(100)$.
In that case, all the trapping processes are done within
$|\phi|<\Lambda$. 
As will be seen in the next section, the ESP in a realistic model of 
anomalous $U(1)$ GUT can have the number of $\chi$ species more than
$\mathcal{O}(100)$.

Our numerical calculation shows that trapping within $|\phi|<\Lambda$
is possible after several oscillations.
This is much less than $\mathcal{O}(100)$, which is naively expected from the first event
of particle production.
Efficient particle production after
the first production is due to the statistical property of the bosonic field. 
For instance, one can estimate that the production rate of the bosonic field
is proportional to $(1+f_B)$, while for the fermionic field the rate is
proportional to $(1-f_F)$.
Here, $f_B$ and $f_F$ are the phase space densities for the bosonic
field and the fermionic field, respectively.
Our numerical calculation shows that $f_B$ grows as $f_B=2.8$,
$17$, $150$,... (for $n=2$ ) and $f_B=5.5$, $115$,...(for $n=3$) at zero momentum. 
Then $f_B$ grows up to ${\cal O}(4000)$ ($n=2$) and 
${\cal O}(20000)$($n=3$), which is enough to explain the enhancement of particle production.

\section{Application to anomalous $U(1)$ GUT}
In this section we consider the possibility of particle production
 and trapping in a phenomenologically viable model, 
 which is called anomalous $U(1)$ grand unified theory
 (GUT)~\cite{Maekawa:2001uk, Bando:2001bj, Maekawa:2001vt,
 Maekawa:2002bk, Maekawa:2002mx, Maekawa:2003bb, Kim:2008kw,
 Ishiduki:2009vr, Kawase:2010na, Maekawa:2012fx}.  
The benefit of this scenario is that one can solve the doublet-triplet
 splitting problem as well as obtaining realistic quark (lepton) masses and
their mixings under the natural assumption that all the interactions
allowed by the symmetry have $\mathcal{O}(1)$ coefficients.
However, since the theory will have an indefinite number of higher
dimensional interactions, there could be an uncountable number of vacua in
the low-energy effective action in general~\cite{Bando:2001bj,Kim:2008kw}.
In this section we will discuss the selection of the vacuum in the light
of trapping. The essential point is that the physically viable vacuum is
near the origin, where all fields have vanishing VEVs and almost all fields
become massless, while the other unphysical vacua are far away from the origin.

\subsection{Toy model}
\label{sec:toy_model}
First, we consider a toy model in which we have the anomalous $U(1)$ gauge symmetry
and the Fayet-Iliopoulos (FI) term.
The superfields $Z_i$  ($i=1,\cdots,n$) have the anomalous $U(1)$
charges $z_i$, which are integer.
The superpotential $W(Z_i)$ contains all the possible interactions
including higher dimensional interaction.
We assume that coefficients of these interactions are always ${\cal O}(1)$.
Here we suppose that they do not have explicit mass terms at the tree level.
\footnote{
In other words, we are considering the effective theory after intetrating out the
superheavy fields with masses $\mathcal{O}(\Lambda)$.
}
The supersymmetric vacuum is determined by the F and D term
conditions:
\begin{equation}
F_{Z_i}^*=-\frac{\partial W}{\partial Z_i}=0,\quad 
D_A=g_A(\xi^2+z_i|Z_i|^2)=0,
\end{equation}
where $\xi^2$ is the Fayet-Iliopoulos parameter ($\Lambda^2 \gg \xi^2 > 0$)
and $g_A$ is the gauge 
coupling constant. Since the potential is given by
\begin{equation}
V_0=\sum_i |F_{Z_i}|^2+\frac{1}{2}D_A^2,
\end{equation}
$F$ and $D$ term conditions will give the minimum of the potential
at $V_0=0$.

The number of the independent F-term conditions is $n-1$, since the
gauge invariance of the superpotential gives an additional relation
$\frac{\partial W}{\partial Z_i}\delta Z_i=0$, where  $\delta Z_i=iz_iZ_i$.
Then, $n-1$ of the $F$-term conditions (complex) and one $D$-term condition (real)
will determine the vacuum of $n$ complex $Z_i$, except for one degree of
gauge transformation (real).
Since the number of the interactions is indefinitely large, there could
be a huge amount of SUSY vacua
that can satisfy $\langle Z_i\rangle= \mathcal{O}(\Lambda)$. 
However, if the number of the fields with a positive charge
($Z^+_i$, $i=1,2,\cdots,n_+$) is
smaller than the number of the fields with a negative charge ($Z^-_i$,
$i=1,2,\cdots,n_-$; $n_+<n_-$), there could be a vacuum with
$\langle Z^+_i\rangle=0$ for all $Z_i^+$.   
In that case, since all the $F$-term conditions of $Z_j^-$ become
trivial, the $F$-term conditions of $Z_i^+$ and the $D$-term condition
will determine the VEVs of $Z_j^-$.
In that case the VEVs of  $Z_j^-$ are much smaller than $\Lambda$, since we have the relation
$\xi^2-|z_i||Z_i|^2=0$ and $\xi^2 \ll \Lambda^2$. 
This solution meets the requirement for the physically interesting vacuum. Especially
when $n_+=n_--1$, the VEVs can be determined by their charges as
\begin{equation}
\langle Z_i\rangle=\left\{
\begin{array}{ll}
 0  & (z_i>0) \\
 \lambda^{-z_i}\Lambda & (z_i\leq 0)
\end{array}\right.,
\label{SUSYvacuum}
\end{equation}
where $\lambda\equiv \xi/\Lambda$. 
This vacuum is quite 
close to the origin where all fields have vanishing VEVs because 
$\langle Z_i\rangle\ll \Lambda$.
The above solution is interesting, in the sense that the coefficients of the
interactions in the effective theory are determined by their $U(1)$ charges.
For example, consider a mass term that is generated from $\lambda^{x+y}\Lambda XY$,
where $x$ and $y$ are the anomalous $U(1)$ charges of $X$ and $Y$,
respectively.
Here $x+y\geq 0$ is assumed.
Note that the mass in the above scenario is generated from the VEVs of
the negatively charged fields $\langle Z_j^-\rangle$.
When $x+y<0$, such mass term is
 forbidden, because $\langle Z_j^+\rangle=0$.
This is called the SUSY zero mechanism. 
In the same way, even for higher dimensional interactions
the effective coefficients can be determined by their charges.
This feature plays an important role in obtaining the realistic
quark and lepton masses and mixings and in solving the doublet-triplet splitting
problem in the realistic models.
Note that in this model the effective mass of the low-energy effective action is
induced by higher dimensional interaction, which explains the
hierarchy of the mass scales.
   
Obviously, the most beautiful and therefore the most attractive point
would be the origin, where the VEVs of all (moduli/Higgs) fields $Z_i$
vanish and a lot of $\chi$ species $Z_i$ become massless.
We call this special point the most enhanced symmetry point (MESP).
Note that $Z_i$ fields behave not only as the moduli fields but also 
as the $\chi$ fields in this model.
(Strictly speaking, the scalar components have masses which are proportional to
the SUSY breaking scale $\xi^2$ and their $U(1)$ charges at the MESP.)
Once all moduli happen to meet together at the MESP at the same time,
they can be trapped at the MESP.  
\footnote{For simplicity, all the fields that are responsible for
quantum particle production are called ``moduli'' when their kinetic
energy is significant, even though their potential could not be negligible.
Also, all the fields that are produced by the quantum effect are
labeled as $\chi$.}
Since it seems unlikely to happen, 
a more plausible process for trapping is that the fields are trapped by
 a step-by-step process. 
We have a lot of attractive hypersurface where one or two  $\chi$ fields become massless.
Once moduli fields pass through the attractive hypersurface, the $\chi$ particles
are produced. Because of the produced effective potential, the moduli fields can 
be trapped on the hypersurface. 
In that way, all moduli fields may be trapped one-by-one.
Note that the supersymmetry is broken at the MESP, where
$V_0=\frac{g_A^2}{2}\xi^4$.
After trapping, as the number densities of the particles are diluted
by the expansion of the Universe, these moduli fields must 
move toward the SUSY vacuum of Eq. (\ref{SUSYvacuum}), which is the nearest
SUSY vacuum from the MESP. The other unphysical vacua with 
$\langle Z_i\rangle \sim \mathcal{O}(\Lambda)$ are far from the MESP.
As a result, the vacuum in eq. (\ref{SUSYvacuum}), which is important to obtain
physically viable GUT,
is selected by the trapping mechanism.
Of course, it is not clear whether such trapping process truly happens or not.
Even if trapping at the MESP does not happen, the vacuum of 
Eq. (\ref{SUSYvacuum}) has much more advantage than the other unphysical vacua
 with $\langle Z_i\rangle\sim \mathcal{O}(\Lambda)$ because of the effective potential induced
 by the produced $\chi$ particles.

\subsection{Realistic models}
There are several realistic SUSY GUT models with anomalous $U(1)$ gauge symmetry.
For example, in $SO(10)$ models, we introduce three $\bf 16$ and one $\bf 10$
in the matter sector and two $\bf 45$, two pairs of $\bf 16$ and $\bf\overline{16}$,
two $\bf 10$, and several singlets in the Higgs sector, and one $\bf 45$
for the vector 
multiplets~\cite{Maekawa:2001uk,Maekawa:2001vt}.
Therefore, more than 1000 real species will be massless at the MESP,
including their superpartners. 
In the simplest $E_6$ model, we introduce three $\bf 27$ in the matter sector, 
two $\bf 78$, two pairs of $\bf 27$ and $\bf\overline{27}$, and several 
singlets in the Higgs sector, 
and one $\bf 78$ for vector multiplets~\cite{Maekawa:2003bb}. 
There are some $E_6$ models with an
additional
pair of $\bf 27$ and $\bf\overline{27}$~\cite{Bando:2001bj,Maekawa:2002bk}. 
They have nearly 2000 massless fields at the MESP. 
Therefore, there are plenty of $\chi$ fields that may cause trapping.
Since in these models the mass terms in the low-energy effective theory
are generated by higher dimensional interaction generically, we are expecting 
that higher dimensional interaction is really responsible for
trapping and vacuum selection.
As we mentioned in the previous subsection, it seems unlikely that 
all moduli fields meet together accidentally at the MESP even once.
We are expecting that fields will be trapped one after another
and finally all moduli fields are trapped around the MESP.

One might be anxious about the decay of
$\chi$ fields~\cite{Kofman:1997yn,Felder:1998vq}, which has not been mentioned in
our analysis. 
Consider the situation in which several moduli fields are already trapped 
on certain attractive hypersurface. 
If those moduli fields are responsible for the mass of some $\chi$ fields 
($\chi_l$), 
basically $\chi_l$ are light. 
When another moduli field comes across other attractive hypersurface,
the $\chi_h$ particles produced by the
moduli field can become heavier than $\chi_l$ at least 
when the moduli field moves away from the attractive hypersurface. 
In that case the heavy field $\chi_h$ could decay into $\chi_l$,
if the lifetime of $\chi_h$ is shorter than the
time scale of the oscillation.
If $\chi_h$ decays into $\chi_l$, the number density of $\chi_h$ becomes
smaller than that without decay, 
and therefore, attractive force to the attractive hypersurface is weakened.
Namely, $\chi_h$ fields with longer lifetime produce more attractive force
to the attractive hypersurface.
It is reasonable to expect that the $\chi$ particles with masses from higher dimensional
interaction have a longer lifetime, because they are lighter and expected to have
higher dimensional interaction even for the decay.
Therefore, such $\chi$ particles may have some advantages for trapping in this situation.
Let us consider another situation in which two moduli fields $A$ and $B$ oscillate 
around different hypersurfaces. When the modulus $A$
($B$) passes through each attractive hypersurface, 
the particles $\chi_A$ ($\chi_B$) are produced. The masses of these $\chi$
particles are
dependent on time because the distance from the hypersurface changes. 
When $\chi_A$ becomes heavier than $\chi_B$, $\chi_A$ can decay into
$\chi_B$, and when $\chi_B$ becomes heavier than $\chi_A$, $\chi_B$ can decay to
$\chi_A$. Therefore, when the moduli field passes through the attractive 
hypersurfaces,
the $\chi$ particles are produced not only by breaking the adiabatic conditions but
also by decaying from other heavy particles. The situation becomes highly complex.
Again, the $\chi$ particles with longer
lifetime have an advantage for trapping.
It is unclear whether the vacuum is trapped at the MESP in the last or not, but
in this subsection, we assume that the vacuum is trapped at least near the
MESP.

The true trapped point may not be the MESP if the masses of fields $\chi$ which 
dominate the energy density are from higher dimensional 
interaction with $n\geq 3$. 
This is because their contributions to the effective potential
cannot dominate $V_0$ around the MESP. 
This may be important to apply the GUT models
because monopoles do not appear in this situation.
(Note that the gauge symmetry is not recovered by trapping.)

Note also that in the above scenario SUSY is broken at the MESP, and therefore,
it is not the true vacuum.
Moreover, because of the SUSY breaking, the scalar component fields
have masses which are proportional to the FI parameter $\xi^2$ and the anomalous 
$U(1)$ charges. The negatively charged fields like GUT Higgs fields have
negative mass square, and therefore they are unstable without the support by 
the produced $\chi$ particles.  
Therefore, the GUT Higgs fields trapped around the MESP are supposed to move away
from the MESP to find the true vacuum after dilution of the produced
$\chi$ fields. 
Basically, the vacuum which satisfies the relation (\ref{SUSYvacuum})
is selected in the last because the other vacua with the VEVs $\mathcal{O}(\Lambda)$
are far from the MESP as discussed in the previous subsection.
\footnote{
The matters cannot have non-vanishing VEVs under Eq. (\ref{SUSYvacuum}),
because matter fields have positive anomalous $U(1)$ charges in the models.
}

However, in the realistic GUT models, we have 
several vacua which satisfy the relations (\ref{SUSYvacuum}).
One of these vacua gives the realistic low-energy effective action.
The situation is very similar to the minimal
SUSY $SU(5)$ GUT, which has three supersymmetric vacua with the gauge 
symmetries $SU(5)$, $SU(4)\times U(1)$, and $SU(3)\times SU(2)\times U(1)$. 
We have not understood why $SU(3)\times SU(2)\times U(1)$ is selected.
Usually, it is expected to happen accidentally.
However, if we study the trapping process and the moving process after trapping
in detail, this important question may be answered.

Finally, we would like to mention the possibility of inflation after trapping.
Our optimistic expectation is that inflation and vacuum selection are both
explained by particle production.
Here the MESP is a local maximum in the effective potential, 
whose height is $\xi^4$.
$\xi$ is just below the usual GUT scale ($\Lambda_{GG}\sim 2\times 10^{16}$),
because of the relation 
$\xi=\lambda\Lambda\sim {\cal O}(1)\times\Lambda_{GG}$~\cite{Maekawa:2001vt,Maekawa:2002mx}. 
More specifically, we can take $\lambda\sim 0.22$ that gives 
$\xi\sim 5\times 10^{15}$ GeV. 
The potential $V_0=(\xi^2-Z)^2$ is too steep to satisfy the slow roll
condition for sufficient $N$ value in usual way.
There are many inflationary models in which trapping plays a crucial
role~\cite{trapping-inflation} and also the many-field model may
accommodate slow-roll inflation on the steep
potential~\cite{Battefeld:2010sw, GUTwarminflation}, but it is unclear if the GUT
theory could be responsible for inflation~\cite{GUTinflation}.
We expect that inflaton becomes sufficiently slow because of
particle production. Since the origin is tremendously
attractive (i.e., quite many massless fields appear at the origin) in
the anomalous $U(1)$ GUT models, the dissipation effect due to particle
production cannot be negligible.
We will study this subject in future.

\section{Summary}
In this paper, we have studied quantum particle production and trapping 
through higher dimensional interaction.
Quantum particle production is possible when the moduli field $\phi$ 
passes through
the non-adiabatic area around the ESP.
We have extended the model~\cite{Kofman:2004yc} and considered
quantum particle production in which the mass of the particles comes
from higher dimensional interaction.
We found that particle production is possible in the non-adiabatic
area, which could be larger than the conventional scenario.
We have confirmed quantum particle production and trapping for
higher dimensional interaction.

Then we have considered the possibility of vacuum selection 
via particle production in realistic GUT models,
which are called anomalous $U(1)$ GUT. 
In such models, the most serious problem called the doublet-triplet splitting can be solved
as well as realistic quark and lepton masses and mixings are obtained
under the reasonable assumption that all interactions including higher dimensional ones
allowed by the symmetry are introduced with $\mathcal{O}(1)$ coefficients.
However, they have infinite unexpected vacua in general because of the natural assumption. 
This issue can be solved by considering vacuum selection by quantum
particle production, because the most attractive point is near the physically viable
vacuum in these models.

It is important to understand why vacuum selection is possible among
various vacua in SUSY models.
Our expectation is that the dynamical history of the Universe, including
quantum creation of the particles and trapping, is important for
that purpose.
Hopefully, our investigation is helpful in understanding the dynamics.

\section{Acknowledgement}
S.E. is supported in part  by the Polish NCN grant DEC-2012/04/A/ST2/00099.
N.M. is supported in part by Grants-in-Aid for Scientific Research from MEXT of 
Japan. This work was partially supported by the Grand-in-Aid for Nagoya University
Leadership Development Program for Space Exploration and Research Program from the MEXT 
of Japan.
T.M wishes to thank his colleagues at Nagoya university.
All authors wish to thank their colleagues at Lancaster
University for their kind hospitality and many invaluable discussions.

\appendix

\section{Calculation of the Number Density} 
\label{sec:calculation_of_produced_particle_number}
In this section we show the analytical estimation of the number density
 $n_\chi$.
The calculation is partly tracing the Chung's method in Ref.~\cite{Chung:1998bt}.

We start with the WKB-type solution of the wave function for $\chi$:
\begin{equation}
 u_k(t) = \frac{\alpha_k(t)}{\sqrt{2\omega_k}} e^{-i\int_{-\infty}^t dt'\omega_k(t')}
   + \frac{\beta_k(t)}{\sqrt{2\omega_k}} e^{+i\int_{-\infty}^t
   dt'\omega_k(t')}, 
\end{equation}
which gives the solution of Eq.(\ref{eq:EOM_of_u}).
Here $\omega_k$ is the frequency defined in Eq.(\ref{eq:frequency}).
The equation of motion for $u_k(t)$ gives 
\begin{eqnarray}
 \dot{\alpha}_k &=& \beta_k \frac{\dot{\omega}_k}{2\omega_k} e^{+2i\int_{-\infty}^t dt' \omega_k},\\
 \dot{\beta}_k &=& \alpha_k \frac{\dot{\omega}_k}{2\omega_k}
  e^{-2i\int_{-\infty}^t dt' \omega_k}. 
\end{eqnarray}
What we need for the calculation is $\beta_k$, since the occupation number can be
evaluated as $n_k = |\beta_k|^2$.

We consider the initial condition $\alpha_k(-\infty)=1,
\beta_k(-\infty)=0$, which means that Eq.(\ref{eq:Bogoliubov_transformed_solution})
is equivalent to Eq.(\ref{eq:WKB_solution_of_u}) at $t=-\infty$. 
 Therefore the zeroth order of the Bogoliubov coefficients can be taken as
$\alpha_k^{(0)}(t)=1, \beta_k^{(0)}(t)=0$.  Using these values, we can obtain the solution of $\beta_k$ as
\begin{equation}
 \beta_k(+\infty) = \int_{-\infty}^{+\infty} dt \: \frac{\dot{\omega}_k(t)}{2\omega_k(t)}
  \exp{\left[ -2i \int_{-\infty}^t dt' \omega_k(t') \right]} \label{eq:solution_of_beta}
\end{equation}
to the leading order.  
Using the zeroth order solution (\ref{eq:WKB_solution_of_phi}), we find the frequency
\begin{equation}
 \omega_k(t) = \sqrt{\mathbf{k}^2 + f(t)},
\end{equation}
where
\begin{equation}
 f(t) \equiv \frac{g^2}{\Lambda^{2(n-1)}} (v^2 t^2 + \mu^2)^n.
\end{equation}
In order to evaluate the integral (\ref{eq:solution_of_beta}), we consider the steepest descent method.
On the complex $t$-plane, there are $2n$ points where $\omega_k(t)=0$ is
satisfied.
These are poles of an integrand in Eq.(\ref{eq:solution_of_beta}).
If we take a path of the integration to go around the lower half,
we need to take into account the poles with the negative imaginary, whose number
is $n$.

For the integer $m$ satisfying $0 \leq m \leq n-1$, we find the poles of
the negative imaginary gives
\begin{equation}
 \tilde{t}_m = \frac{1}{v} \sqrt{ -\mu^2 + \left( \frac{\Lambda^{n-1}}{g} \mathbf{|k|} \right)^{\frac{2}{n}}
  e^{-i\frac{2m+1}{n}}}. \label{eq:saddle_point}
\end{equation}

\begin{figure}[t]
 \begin{center}
  \includegraphics[scale=0.8]{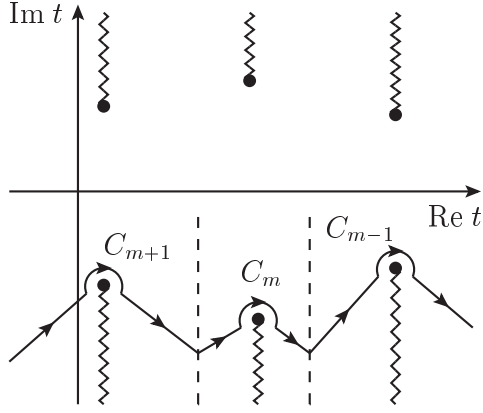}
  \caption{The path of integration on the $t$ complex plane.
  Dots show the poles, and wavy lines show the branch cuts.}
  \label{fig:path}
 \end{center}
\end{figure}

The exponential of (\ref{eq:solution_of_beta}) around $\tilde{t}_m$
can be expanded as
\begin{eqnarray}
 \int_{-\infty}^t dt' \omega_k(t') &=& \int_{-\infty}^{\tilde{t}_m} dt' \omega_k(t') \nonumber \\
  & & + \int_{\tilde{t}_m}^t dt' \sqrt{\mathbf{k}^2 + f(\tilde{t}_m) + \dot{f}(\tilde{t}_m)(t'-\tilde{t}_m)+\dots}\\
 &=& \int_{-\infty}^{\tilde{t}_m} dt' \omega_k(t') + \frac{2}{3} \sqrt{\dot{f}(\tilde{t}_m)} \: (t - \tilde{t}_m)^{3/2}
  + \dots \label{eq:expansion_of_integral_of_beta}
\end{eqnarray}
where we have used the relation $\mathbf{k}^2 + f(\tilde{t}_m)=0$.  Using (\ref{eq:expansion_of_integral_of_beta}) and
\begin{eqnarray}
 \frac{\dot{\omega}_k}{2\omega_k}
  &=& \frac{1}{4} \: \frac{\dot{f}(\tilde{t}) + \dots}{\mathbf{k}^2 + f(\tilde{t}_m)
   + \dot{f}(\tilde{t}_m) (t-\tilde{t}_m) + \dots} \nonumber \\
  &=& \frac{1}{4(t-\tilde{t}_m)} + \dots,
\end{eqnarray}
we find that Eq.(\ref{eq:solution_of_beta}) can be rewritten as
\begin{equation}
 \beta_k \sim \sum_m U_m \exp{\left[ -2i \int_{-\infty}^{\tilde{t}_m} dt' \omega_k(t') \right]}, \label{eq:beta2}
\end{equation}
where
\begin{equation}
  U_m \equiv \frac{1}{4} \int_{C_m} \frac{dt}{t-\tilde{t}_m}
   \exp{\left[-i\frac{4}{3} \sqrt{\dot{f}(\tilde{t}_m)} \: (t-\tilde{t}_m)^{3/2} \right]}. \label{eq:U}
\end{equation}
$C_m$ denotes the path which approaches a pole on $t=\tilde{t}_m$
along the steepest descent path and goes around the pole (and leaves the pole).
(See Figure \ref{fig:path}).
If we take the angle going around the pole to be $4\pi/3$, the
outward-goind path becomes the steepest descent path.
Then, we can obtain
\begin{equation}
 U_m \sim \frac{i\pi}{3}. \label{eq:U2}
\end{equation}
Using the above method, the occupation number can be calculated as
\begin{eqnarray}
 n_k &=& |\beta_k|^2\\
  &\sim& \frac{\pi^2}{9} \sum_{m,m'}
   \exp{\left[ -2i \left\{ \left( \int_{-\infty}^{\tilde{t}_m}dt \omega_k \right)
   - \left( \int_{-\infty}^{\tilde{t}_{m'}}dt \omega_k \right)^* \right\} \right]}, \label{eq:occupation_number}
\end{eqnarray}
which can be evaluated by separating the path of the integration $(-\infty, \tilde{t}_m)$ into two parts:
\begin{equation}
 \int_{-\infty}^{\tilde{t}_m} dt' \omega_k(t') = \Phi_m + \Omega_m,
\end{equation}
where
\begin{eqnarray}
 \Phi_m &\equiv& \int_{-\infty}^0 dt' \omega_k(t'), \label{eq:Phi} \\
 \Omega_m &\equiv& \int_0^{\tilde{t}_m} dt' \omega_k(t'). \label{eq:Omega}
\end{eqnarray}
In Eq.(\ref{eq:Phi}), we considered the path that approaches $t=0$ along the real
axis.
In Eq.(\ref{eq:Omega}), we considered the path going toward the pole ($\tilde{t}_m$)
with the fixed phase of the complex time.
Since $\Phi$ is real, (\ref{eq:occupation_number}) can be rewritten as
\begin{equation}
 n_k \sim \frac{\pi^2}{9} \sum_{m,m'}
  \exp{\left[ -2i \left( \Omega_m - \Omega_{m'}^* \right) \right]}. \label{eq:occupation_number2}
\end{equation}
Note that in the above calculation $\Phi_m$ is not important.

Calculating the integration of Eq.(\ref{eq:occupation_number2}), we find
\begin{equation}
 n_{\chi} \sim \int \frac{d^3k}{(2\pi)^3} \frac{\pi^2}{9} \sum_{m,m'}
  \exp{\left[ -2i \left( \Omega_m - \Omega_{m'}^* \right) \right]}. \label{eq:number_density}
\end{equation}
This integral can be performed for $n=1$, which reproduces the
conventional result
\begin{equation}
 n_{\chi} \sim \frac{\pi^2}{9} \frac{(gv)^{3/2}}{(2\pi)^3} e^{ - \pi g\mu^2/v}.
\end{equation}
Therefore, the approximation gives the estimation that is
$\pi^2/9 \sim 1.10$ times larger than
the exact result in Eq.(\ref{eq:number_density_for_n=1}).

For $n \geq 2$, it is difficult to calculate the integral in Eq.(\ref{eq:number_density}).
Exceptionally, when $\phi$ hits on the ESP ($\mu=0$), one can calculate
the number density $n_{\chi}$ as
\begin{equation}
 \left. n_{\chi} \right|_{\mu=0} = \frac{1}{18} \frac{n}{n+1} (g\Lambda)^3 A_n^{-\frac{3n}{n+1}}
  \Gamma \left( \frac{3n}{n+1} \right) \sum_{m,m'} \left[ \sin{\frac{m+m'+1}{2n}\pi} \right]^{-\frac{3n}{n+1}}
   \cos{\frac{3}{2} \frac{m-m'}{n+1}\pi}, \label{eq:n_chi_0th}
\end{equation}
where
\begin{equation}
 A_n \equiv 2 B \left( 1+\frac{1}{2n}, \frac{1}{2} \right) \cdot \frac{g\Lambda^2}{v}.
\end{equation}
Here $B(\alpha, \beta)$ is a beta function defined by
\begin{equation}
 B(\alpha, \beta) \equiv \int_0^1 dx x^{\alpha-1} (1-x)^{\beta-1}
  = \frac{\Gamma \left( \alpha \right)\Gamma \left( \beta \right)}{\Gamma \left( \alpha + \beta \right)}.
\end{equation}
In the above calculation we considered 
\begin{equation}
 g^{1/n}v/\Lambda^2 \ll 1
\end{equation}

Let us extend the above calculation from $\mu=0$ to $\mu \neq 0$.
Considering $\mu\ll \Lambda$, we can evaluate (\ref{eq:number_density})
by expanding the equation around $\mu=0$, which shows 
\begin{equation}
 n_{\chi} = \left. n_{\chi} \right|_{\mu=0} + \mu^2 \cdot \left. \frac{dn_{\chi}}{d\mu^2} \right|_{\mu=0}
  + \mathcal{O} \left( \mu^4 \right).
\end{equation}
As mentioned above, the first term gives Eq.(\ref{eq:n_chi_0th}).
We find that the second term gives
\begin{eqnarray}
 \mu^2 \cdot \left. \frac{dn_{\chi}}{d\mu^2} \right|_{\mu=0} &=&
  \mu^2 \cdot \frac{\pi^2}{9} \int \frac{d^3k}{(2\pi)^3} \sum_{m,m'} e^{-2i\left(\Omega_m-\Omega_{m'}^*\right)}
   (-2i) \left. \left( \frac{d\Omega_m}{d\mu^2} - \frac{d\Omega_{m'}^*}{d\mu^2} \right) \right|_{\mu=0}\\
 &=& -\frac{1}{2} \frac{\mu^2}{\Lambda^2} \cdot \frac{1}{18} \frac{n}{n+1} (g\Lambda)^3 A_n^{-\frac{3n-2}{n+1}}
  \Gamma \left( \frac{4n-1}{n+1} \right)
   \frac{B \left( 1-\frac{1}{2n}, \frac{1}{2} \right)}{B \left( 1+\frac{1}{2n}, \frac{1}{2} \right)} \nonumber \\
  & & \times \sum_{m,m'} \left[ \sin{\frac{m+m'+1}{2n}\pi} \right]^{-\frac{3n-2}{n+1}} \cos{\frac{5}{2} \frac{m-m'}{n+1}\pi}. \label{eq:n_chi_1th}
\end{eqnarray}

Finally, we found the number density near the ESP as

$\quad \bullet \: n=1$ :
\begin{equation}
 n_{\chi} \sim \frac{\pi^2}{9} \frac{(gv)^{3/2}}{(2\pi)^3} e^{ - \pi g\mu^2/v}.
 \label{nchi_n1}
\end{equation}

$\quad \bullet \: n \geq 2$ :
\begin{eqnarray}
 n_{\chi} & \sim & \frac{1}{18} \frac{n}{n+1} (g\Lambda)^3 A_n^{-\frac{3n}{n+1}} \Gamma \left( \frac{3n}{n+1} \right) \nonumber \\
 & & \times \sum_{m,m'} \left\{ \left[ \sin{\frac{m+m'+1}{2n}\pi} \right]^{-\frac{3n}{n+1}} \cos{\frac{3}{2} \frac{m-m'}{n+1}\pi} \right. \nonumber \\
 & & - \frac{1}{2} \frac{\mu^2}{\Lambda^2} \cdot A_n^{\frac{2}{n+1}}
  \frac{\Gamma \left( \frac{4n-1}{n+1} \right)}{\Gamma \left( \frac{3n}{n+1} \right)}
  \frac{B \left( 1-\frac{1}{2n}, \frac{1}{2} \right)}{B \left( 1+\frac{1}{2n}, \frac{1}{2} \right)}
  \left[ \sin{\frac{m+m'+1}{2n}\pi} \right]^{-\frac{3n-2}{n+1}} \cos{\frac{5}{2} \frac{m-m'}{n+1}\pi} \nonumber \\
 & & \left. + \mathcal{O} \left( \mu^4 \right) \right\} \label{eq:n_chi_final} \\
 & \equiv & C_n^{(0)} \cdot (gv)^\frac{3}{2} \left( \frac{v}{g\Lambda^2} \right)^{\frac{3(n-1)}{2(n+1)}}
  \left( 1 + C_n^{(1)} \cdot \left( \frac{g\Lambda^2}{v} \right)^{\frac{2}{n+1}} \frac{\mu^2}{\Lambda^2}
   + \mathcal{O} \left( \mu^4 \right) \right).\label{eq:number_density_result}
\end{eqnarray}
In the final line, we defined coefficients $C_n^{(0)}, C_n^{(1)}$ as
\begin{eqnarray}
 C_n^{(0)} & \equiv & \frac{1}{18} \frac{n}{n+1} \left[ 2 B \left( 1+\frac{1}{2n}, \frac{1}{2} \right) \right]^{-\frac{3n}{n+1}}
  \Gamma \left( \frac{3n}{n+1} \right) \nonumber \\
  & & \quad \times \sum_{m,m'}
   \left[ \sin{\frac{m+m'+1}{2n}\pi} \right]^{-\frac{3n}{n+1}} \cos{\frac{3}{2} \frac{m-m'}{n+1}\pi},\\
 C_n^{(1)} & \equiv & -\frac{1}{2}\left[ 2 B \left( 1+\frac{1}{2n}, \frac{1}{2} \right) \right]^{\frac{2}{n+1}}
  \frac{\Gamma \left( \frac{4n-1}{n+1} \right)}{\Gamma \left( \frac{3n}{n+1} \right)}
  \frac{B \left( 1-\frac{1}{2n}, \frac{1}{2} \right)}{B \left( 1+\frac{1}{2n}, \frac{1}{2} \right)} \nonumber \\
  & & \quad \times \frac{\sum_{m,m'} \left[ \sin{\frac{m+m'+1}{2n}\pi} \right]^{-\frac{3n-2}{n+1}} \cos{\frac{5}{2} \frac{m-m'}{n+1}\pi}}
   {\sum_{m,m'} \left[ \sin{\frac{m+m'+1}{2n}\pi} \right]^{-\frac{3n}{n+1}} \cos{\frac{3}{2} \frac{m-m'}{n+1}\pi}}.
\end{eqnarray}
These coefficients are determined by $n$.  The explicit values of the coefficients
are given in Table \ref{tab:coefficients} for $n=1,2,3$.

\end{document}